%% file: ferrogels.tex
\documentclass[twocolumn,jcp]{revtex4-1}
\usepackage{graphicx} 
\usepackage{grffile}
\usepackage{pgfplots}
\pgfplotsset{
  /pgfplots/ybar legend/.style = {
	/pgfplots/legend image code/.code={%
	\draw[##1,/tikz/.cd,bar width=3pt,yshift=-.2em,bar shift=0pt]
	plot coordinates {(0cm,0.8em)};}
  }
}
\usepackage{amsmath}

\newcommand{\stdFigWidth}{8.9cm}

\begin{document}
\title{Ferrogels cross-linked by magnetic particles: Field-driven deformation and elasticity studied using computer simulations}

  \author{Rudolf Weeber$^{1}$}\email{weeber@icp.uni-stuttgart.de} 
  \author{Sofia Kantorovich$^{2,3}$}\email{sofia.kantorovich@univie.ac.at}
  \author{Christian Holm$^{1}$}\email{holm@icp.uni-stuttgart.de} 

\affiliation{$^1$
Institute f\"ur Computerphysik,
Universit\"{a}t Stuttgart,
Allmandring 3,
70569 Stuttgart,
Germany
}%
\affiliation{$^2$
Ural Federal University, 
Lenin av. 51, 
620083, Ekaterinburg, 
Russia
}%
\affiliation{$^3$
Universit\"at Wien, Sensengasse 8, 1090, Wien, Austria
}%

\date{\today}

\begin{abstract} 

  Ferrogels, i.e. swollen polymer networks into which magnetic particles are immersed, can be considered as ``smart materials'' since their shape and elasticity can be controlled by an external magnetic field. Using molecular dynamics simulations on the coarse-grained level we study a ferrogel in which the magnetic particles act as the cross-linkers of the polymer network.  In a homogeneous external magnetic field the direct coupling between the orientation of the magnetic moments and the polymers by means of covalent bonds gives rise to a deformation of the gel, independent of the interparticle dipole-dipole interaction.  In this paper we report quantitative measurements of this deformation, and in particular, we investigate the gel's elastic moduli and its magnetic response for two different connectivities of the network nodes. Our results demonstrate that these properties depend significantly on the topology of the polymer network.

\end{abstract}


\maketitle

\section{Introduction}

Polymer gels combined with magnetic nanoparticles are composite soft materials whose elastic properties can be controlled by external magnetic fields \cite{barsi96a}.  In the last decade an active research in this area put forward several terms to characterise these materials: magnetoelastomers, ferrogels magnetic gels, etc.  These materials gained attention both, in fundamental research and applications, as they combine the intrinsic properties of a hydrogel with a relatively high sensitivity to magnetic fields inherent to magnetic fluids \cite{rosensweig85a}. Of particular importance is the potential to change the shape and mechanical characteristics of a ferrogel by applying a relatively weak magnetic field \cite{zrinyi96a,gollwitzer08a,filipcsei10a,wood11a,weeber12a,annunziata13a}.

The ability to control the shape of a magnetic gels gave rise to a number of new applications. Among them are drug delivery \cite{hu07a,qin09b}, cancer treatment strategies (hyperthermia) \cite{babincova01a}, actuation \cite{ramanujan06a,monz08a} and transport \cite{kondo94a,wang07d}.

On a microscopical level, three ingredients need to be considered: the magnetic fluid, the polymer network, and the coupling between them.  First, magnetic fluids (ferrofluids) have been available for decades. They have been studied extensively experimentally and theoretically, and are used in technical and medical applications\cite{odenbach09a,jurgons06a,holm05b}.  The magnetic nanoparticles typically have a size on the order of 10 nm and hence consist of a single ferromagnetic domain. This implies that they carry a permanent magnetic moment.  Due to Brownian motion, the magnetic particles in a ferrofluid are randomly oriented and the fluid as a whole does not exhibit a net magnetic moment. This changes, once an external field is applied, and the magnetic moments in the fluid can align to it.  There are two mechanisms by which this can occur \cite{koetitz95a}: the magnetic moments of the particles can re-align internally, without a rotation of the particle as a whole (N$\mathrm{\acute{e}}$el mechanism), or the particles as a whole can rotate to align their moments (Brownian mechanism).  In the case of magnetic gels, it is of particular importance to know which mechanism is present, because only for the Brownian mechanism a direct coupling between the orientation of the magnetic moments and the polymers is possible. Moreover one can also study ferrogels, where the magnetic nanoparticles have a finite uniaxial magnetic anisotropy \cite{ryzhkov15a}.

Looking at the second ingredient, hydrogels, i.e. water-filled polymer networks, are common in everyday applications such as contact-lenses, food and medical items \cite{peppas00a,chiessi07a,hoffman12a,hoepfner14a}. They can be synthesized from a wide range of polymers and their chemical and physical properties can be tailored to suite the applications. For instance, by varying the degree of cross-linking between the polymers, their stiffness can be controlled. Also, hydrogels can be made sensitive to environmental factors such as temperature and ph-value.
In the context of magnetic gels, it is particularly important to distinguish between the type of cross-linking between the polymers.
This can be either due to comparatively weak interactions such as van der Wals forces and hydrogen bonds, or it can be due to covalent bonds.
While the links between the polymers are dynamically changing in the first case, in the latter case, the bonds are stable and the network topology does not change over time. Simulations of hydrogels have been performed, but the literature as by far not as extensive as for polymer solutions \cite{kosovan13a} 

\begin{figure} 
\begin{center} 
\includegraphics[width=\linewidth]{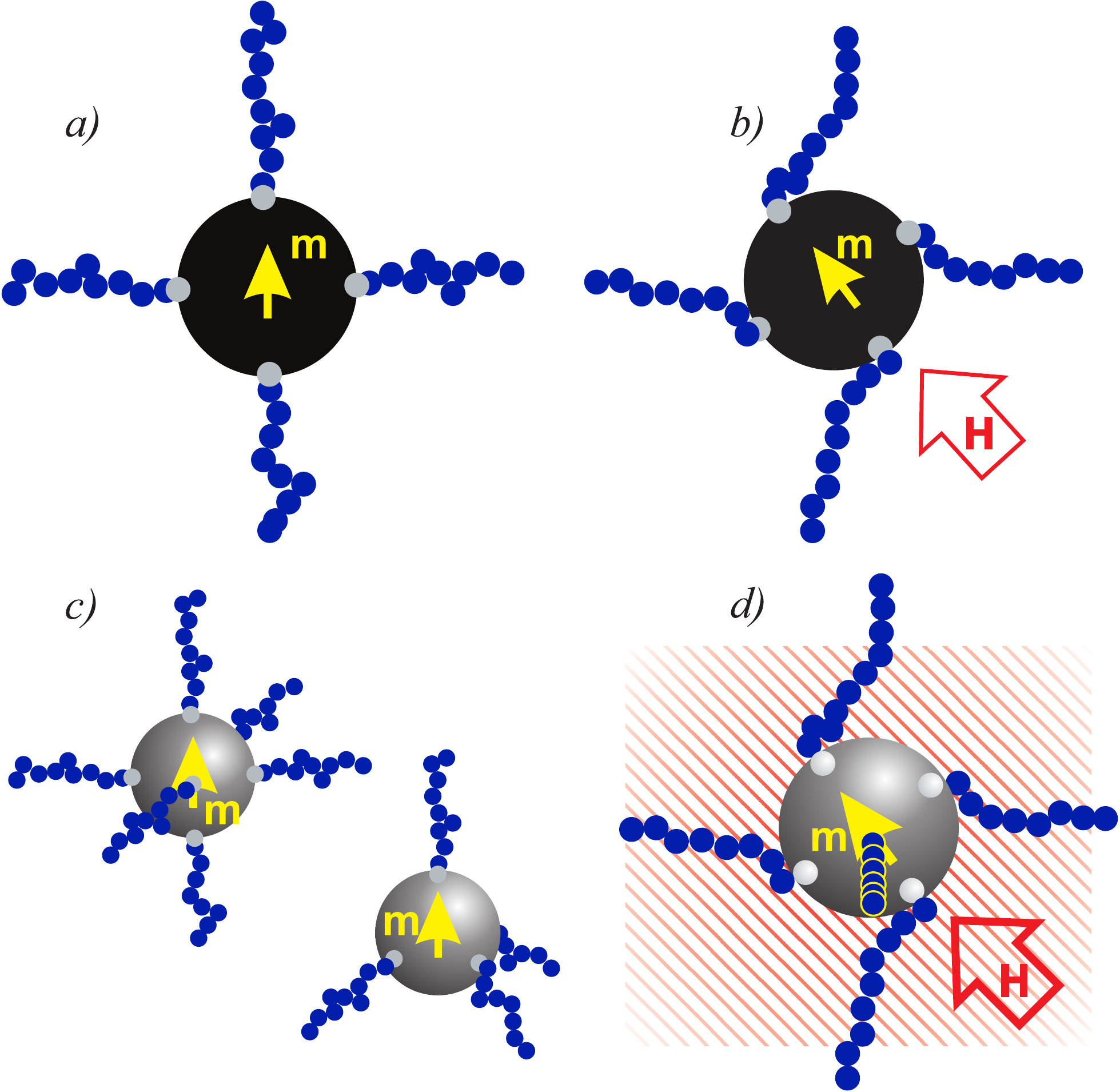} 
\end{center} 
\caption{\label{fig:sketch} Sketch of magnetic nanoparticles in a gel. Upper row (a) and b)): 2D model, left without an applied external field, right with an applied external field, as marked with the red arrow. One can see four polymer chains rigidly bound to specific spots (light blue, virtual sites) on the surface of a magnetic nanoparticle. When a field is applied and the magnetic particle aligns its dipole to it, the polymer chains wrap around the particle. This exerts a stress on the chain, which leads to a shrinking of the ferrogel (isotropic in 2D). Lower row (c) and d)) 3D model, left without an external field, right with an external field applied. In sketch c) a 4-fold (diamond) and 6-fold (cubic) crosslinker is shown. In sketch d), an external magnetic field H is applied within the plane, as shown with the red arrow. Notice that only chains within the plane are subjected to the deformation, the polymer chains perpendicular to the plane are not affected. This leads to an anisotropic shrinking of the gel in 3D.  } 
\end{figure}

The third thing to be considered when studying a ferrogel is the coupling between the polymers and the magnetic particles.  One possibility is a loose coupling by means of hydrogen bonding or van der Wals interactions. In this case, the magnetic particles can exert a stress on the polymer matrix as they move, but the rotational degree of freedom of the nanoparticles is not directly coupled to the polymers.  A rotational coupling between the nanoparticles and the polymers can be achieved by binding the polymers covalently to specific spots on the surface of the magnetic nanoparticles\cite{dudek07a,messing11a,roeder14a}. When the particles rotate due to an external field, the polymers wrap around the particles. This, in turn creates a stress on the polymer chain which leads to a shrinking of the gel. A sketch illustrating this mechanism can be found in Fig.\,\ref{fig:sketch}.  In this case, the magnetic particles act as actual cross-linkers of the network\cite{ilg13a}.

 In this paper, we consider such a magnetic nanoparticle cross-linked gel, which due to the covalent bonds has a network topology which is constant over time. The magnetic moments are assumed to relax via the Brownian mechanism, i.e., by a rotation of the particle as a whole.

Since many applications of ferrogels are connected to actuation in an external magnetic field, it is crucial to understand the microscopic mechanisms by which the gels are deformed.  First, in a field gradient, the magnetic particles tend to accumulate in regions with a stronger field, due to magnetostriction \cite{zrinyi96a}.  Second, in a homogeneous field, a deformation can occur due to a change of the interactions between the magnetic particles as they are aligning in an external field.  The dipole moments in the gel system are randomly oriented in the field-free case, and there is on average only a weak interaction between them, which is not strong enough to overcome the elasticity of the polymer matrix.  The interaction between magnetic particles can become significant, however, as their moments are aligned in a field. Due to this change in interaction, the particles rearrange and thereby deform the matrix.  This is the microscopic equivalent of the fact that a homogeneously magnetized elastic medium elongates in field direction in order to reduce its demagnetization field\cite{raikher03a,zubarev12a,ivaneyko13a}.  This mechanism was observed, e.g., in Ref.\,\cite{gollwitzer08a}.  It is noteworthy that the type of local particle configuration influences the response to a field\cite{ivaneyko11a,stolbov11a}.  Finally, a third deformation mechanism arises from a direct coupling of the orientation of the magnetic moments to the polymers as described above.  A detailed study on the interaction between two magnetic particles connected by a polymer chain in this way can be found in \cite{pessot15a}.  

In our earlier work\cite{weeber12a}, we presented two gel models in two dimensions, which deform by the second and third mechanism, respectively.  For the case of a gel cross-linked by magnetic particles (model II in Ref.\,\cite{weeber12a}), an isotropic shrinking can be observed. This is, because in 2D, there is only one rotation axis, namely the one perpendicular to the model plane. Once the magnetic particle is rotated by the field, all chains attached to it receive the same amount of stress.  In Ref.\,\cite{weeber15a}, however, we have shown that the shrinking is no longer isotropic in three dimensions.  This can be explained by the effect that the rotation always takes place around an axis which is perpendicular to both, the magnetic moment prior to alignment and the direction of the external magnetic field.  Polymer chains that are attached within a plane spanned by these two vectors and that are going through the particle's center follow the full rotation of the magnetic particle.  Chains attached parallel to the rotation axis, on the other hand, are not affected by the particle's rotation.  Prior to the alignment by a field, the magnetic moments of different particles are random, but the rotation axis around which the alignment takes place is always perpendicular to the external field. Hence, there is less deformation in the directions perpendicular to the field.  A sketch explaining this mechanism can be found in Fig.\,\ref{fig:sketch}.

In contrast to our previous works, we present here a detailed quantitative study of the deformation, elasticity and magnetic response of a three-dimensional gel cross-linked by magnetic particles, which has been investigated in depth in the PhD work of the first author\cite{weeber14a}. We consider two topologies of networks, a simple cubic (SC) network with six-fold connectivity and a diamond topology with fourfold connectivity (DC). 
The paper is structured as follows: In Sec.\,\ref{sec:sim}, we describe the simulation technique used, and then introduce the model. 
Thereafter, the equilibrium degree of swelling of the gel is obtained (Sec.\,\ref{sec:eqswelling}).
In Sec.\,\ref{sec:aniso-def} the gel's deformation in a field is measured. Sec.\,\ref{sec:el-const} deals with the measurement of the system's elastic constants and in Sec.\,\ref{sec:mag}, the magnetic response is examined. 
We conclude the paper with a summary.

\section{Simulation technique}
\label{sec:sim}

In this paper, we study the deformation of three-dimensional magnetic gels cross-linked by magnetic nanoparticles in a magnetic field.
This is done by means of molecular dynamics simulations on a coarse-grained level, using the ESPResSo software\cite{limbach06a,arnold13a}: rather than considering individual atoms as degrees of freedom, the smallest units considered in the simulations are the units of polymer that roughly match one Kuhn length , and the magnetic nanoparticles. In this way, the number of degrees of freedom in the system is reduced to a point, at which it is possible to simulate several meshes of the gel in each Cartesian direction.
The simulations are performed in the canonical NVT ensemble using a Langevin thermostat.
For each translational degree of freedom, the equation of motion is
\begin{equation}
\label{eqn:langevin}
m\ddot{x} =-\gamma \dot{x} +F +F_{\rm{random}} ,
\end{equation}
where $m$ denotes the mass, $\gamma$ the friction coefficient, $F$ the force due to other particles, and $F_{\rm{random}}$ a random force mimicking collisions with the surrounding solvent.
According to the fluctuation-dissipation theorem, in order to maintain a thermal energy of $k_B T$, each random force component has to have a mean of zero and a variance of
\begin{equation}
\langle F_{\rm{random}}^2 \rangle =2 k_B T \gamma.
\end{equation}
Rotational degrees of freedom are treated in a similar way. In Eqn.\,\ref{eqn:langevin}, position, mass and force are replaced by orientation, inertial tensor and torque, respectively\cite{wang02a}.

The polymer chains are represented as bead-spring models, in which the beads are bound by harmonic potentials 
\begin{equation}
\label{eqn:harmonic}
U_{\rm{harmonic}}(r) =\frac12 k (r-r_0)^2,
\end{equation}
where $k$ is the spring constant and $r_0$ the equilibrium distance.

Both, the beads forming the polymer chains and the magnetic particles, are represented as soft spheres interacting via a purely repulsive Lennard-Jones interaction, the so-called WCA-potential\cite{weeks71a}. Its functional form is given by
\begin{equation}
\label{eqn:wca}
U_{WCA} = 
\begin{cases}
4\epsilon
\left[\left(\frac{\sigma}{r_{ij}}\right)^{12}
 - \left(\frac{\sigma}{r_{ij}}\right)^{6}\right]+\epsilon & r<r_c\\
0 & r \geq r_c
\end{cases}
\end{equation}
where $\epsilon$ controls the energy scale of the potential, $\sigma$ is the sum of the radii of the interacting beads, and $r_c=2^{1/6}\sigma$ denotes the cut-off distance after which the potential is zero.

The anchoring point of the covalent bonds between the magnetic nanoparticle and the polymer chains attached to it is modelled using the virtual site feature of ESPResSo \cite{arnold13a}:.
Virtual sites are particles whose position is not determined by integrating an equation of motion, rather it is calculated from the position and orientation of other particles.
In this particular case, a virtual site is placed immediately under the surface of the magnetic nanoparticle to form a binding site for the attached polymer. The latter is rigidly connected to the magnetic particle and will follow both, its translational and rotational motion. The technical details can be found in the appendix of Ref.\,\cite{weeber12a} as well as in \cite{arnold13a}.
A sketch of a magnetic particle with the attached polymer chain can be found in Fig.\,\ref{fig:sketch}.

\begin{figure*}
\noindent
\begin{center}
\includegraphics[width=\stdFigWidth]{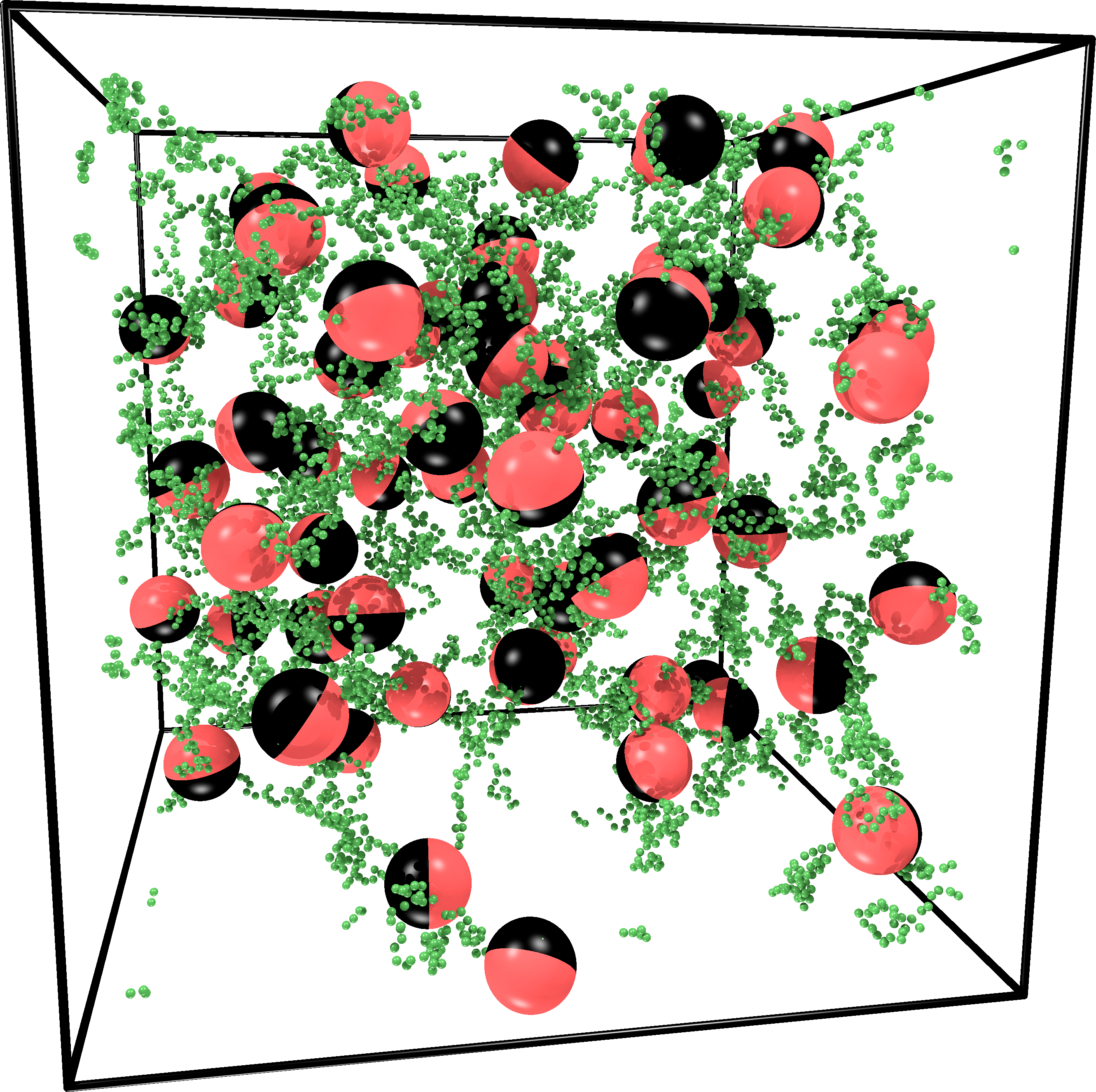}
\includegraphics[width=\stdFigWidth]{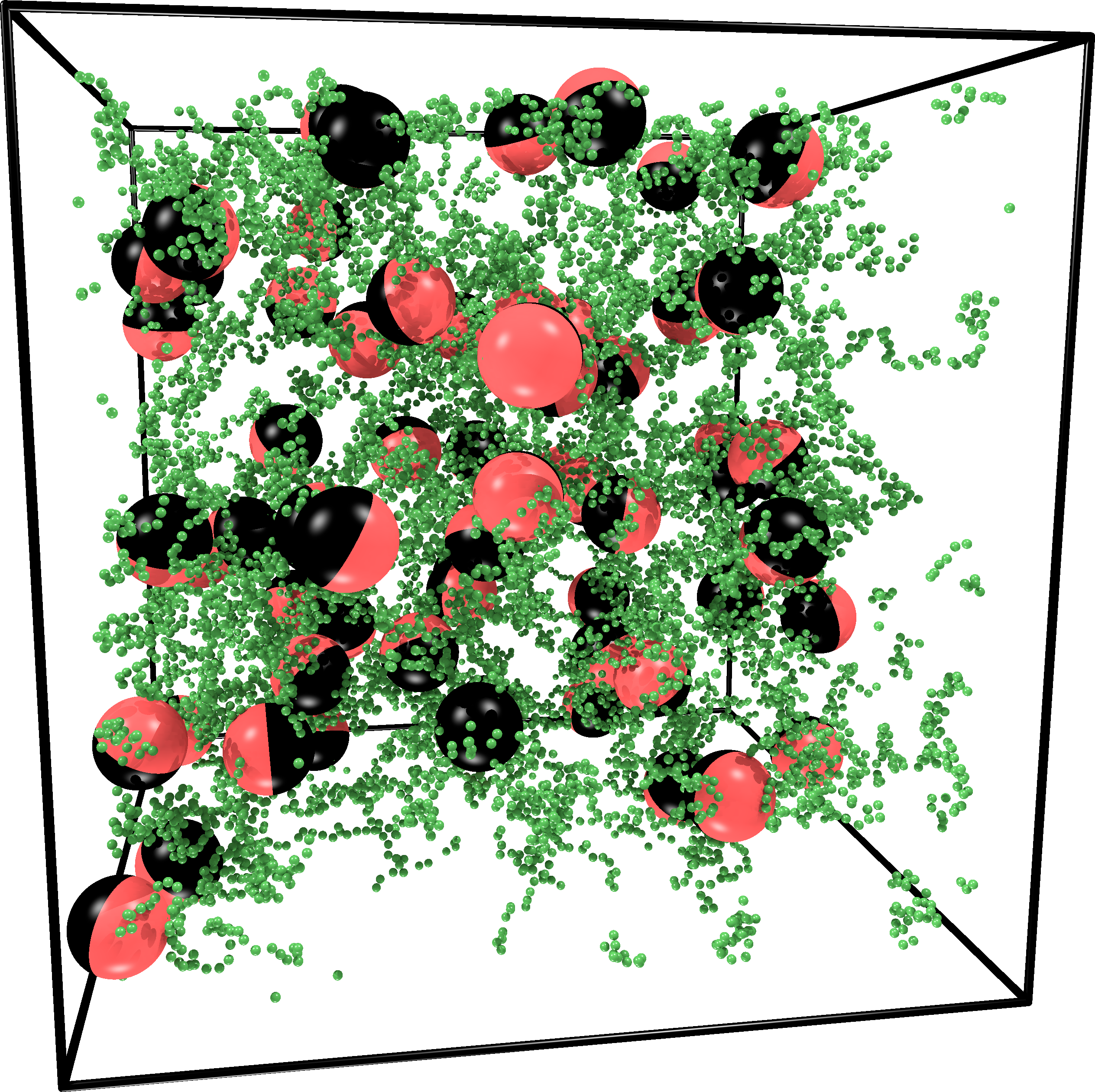}
\end{center}
\caption{\label{fig:fan-3d-system}
Snapshot of a gel sample at equilibrium swelling in the diamond cubic geometry with four chains attached to a node (left) and the simple cubic geometry with six chains attached to a node (right). 
Consisting of 60 beads, the chains are connected to specific spots on the surface of the magnetic particles. In the field-free case, depicted here, the magnetic moments are randomly aligned.
}
\end{figure*}
The magnetic particles, forming the nodes of the network, are initially placed on a regular three-dimensional lattice. Then, the chains are attached to a specific point on the nodes' surface. Therefore, when the node particles rotate, the chain ends have to follow.
Before the chains are connected, the dipole moments of the node particles are randomized, to make the system isotropic.
The system is studied taking into account only the dipole-field interaction, but not the dipole-dipole interaction.
We do this for three reasons: first, the density of magnetic particles is relatively low, so that the influence of the dipole-dipole interaction is rather weak. This was checked in the 2D case discussed in Ref.\,\cite{weeber12a}. Second, the inclusion of the dipole-dipole interaction could result in two deformation mechanisms occurring at the same time, which would make it more difficult to attribute findings to a specific mechanism. 
Third, for particles placed on a regular lattice,
the kind of deformation due to dipole-dipole interactions depends on the lattice structure\cite{ivaneyko12a}.
In order to avoid artefacts, these interactions should therefore only be included in a more randomized system.

In this work, two simple choices for an initial lattice are considered. These are the diamond cubic (DC) and simple cubic (SC) geometries. In the former case, four chain ends are connected to each node, while there are six in the latter case.
By switching from the diamond cubic to the simple cubic structure, one can thus vary the degree of connectivity (and in consequence the elastic and magnetic response) of the network.

In contrast to the 2D case,
which we studied in model II of Ref.\,\cite{weeber12a}, periodic boundary conditions are applied, in order to reduce surface effects, and to study a truly macroscopic gel.
Snapshots of samples after equilibration are shown in Fig.\,\ref{fig:fan-3d-system} for the DC and SC lattice.
The simulation parameters are as follows:
the gel consists of $N_n=64$ node particles as well as $N_c^{\rm{dc}}=2 N_n$ chains in the diamond cubic structure, or $N_c^{\rm{sc}}=3 N_n$ chains in the simple cubic structure.
The chains consist of 60 or 80 beads, with a diameter of $\sigma_c=1$, whereas the node particles have a diameter of $\sigma_n=10$. The scale of the soft sphere interaction (Eqn.\,\ref{eqn:wca}) is $\epsilon=10$.
The bonds between neighboring particles in a chain as well as between the ends of the chains and the virtual sites interact via a harmonic potential (Eqn.\,\ref{eqn:harmonic}) with an equilibrium elongation of $2^{\frac16} \sigma$, coinciding with the cut-off of the WCA-potential. The spring constant of the harmonic potential is $k=10$.

As the shape of the gel for the given parameters is not known a priori, 
an iterative procedure is applied to adjust the box size. 
We begin with an estimated shape. 
Then, iteratively, the stress is measured and the box is shrunk or expanded in each coordinate, according to the diagonal elements of the observed stress tensor. The orthogonal basis of the stress tensor coincides with the orientation of the simulation box.
After the box shape is adjusted, a new simulation is performed. This is repeated, until the measured stress approaches zero with the desired tolerance.
When the gel can be assumed to be isotropic, i.e., when no external field is applied, an optimized procedure can be used, which will be explained in the next section.
Even in the field free case, there will be a small anisotropy in an individual sample of the gel, because the sum of the randomly drawn initial dipole moments is never exactly zero.
The stress is measured and averaged for several instances of the system at any given set of parameters to compensate for this.
In other words, the simulation is divided into the following steps:
\begin{enumerate}
\item The magnetic node particles are placed on the lattice with random dipole orientations.  
The distance between the surfaces of adjacent nodes is such that the chains can be inserted in an elongated configuration.
Hence, the resulting initial lattice constant is 
\begin{equation}
a=\sigma_n +l_c \sigma_c,
\end{equation}
where $\sigma_n$ and $\sigma_c$, are the Lennard Jones diameters of the node and chain particles, and $l_c$ is the number of particles in a chain.
\item Chains are connected to the surface of two adjacent nodes. Four chain ends are connected to a node in the diamond cubic case and six in the simple cubic case. 
\item After the network is cross-linked, it is scaled to the desired shape. This is done by adjusting the box size in 400 steps from the initial shape to the target shape. After each deformation step, the system is equilibrated with the Langevin thermostat for 100 iterations of time step $dt_{\rm{reshape}}=0.0006$ to disperse the energy introduced by the deformation.
\item When the final shape is reached, the system is equilibrated further for 200\,000 steps of size $dt=0.01$, again using the Langevin thermostat.
\item Finally, observables can be measured. The time step is again $dt=0.01$ and stress measurements are taken every 20 time steps, whereas the magnetization is measured every 100\,000 steps. In a typical simulation, more than 50\,000 stress measurements are taken. This enormous amount of statistics is needed due to the large fluctuations of the measured stress due to the softness of the material.
\end{enumerate}

\section{Equilibrium swelling}
\label{sec:eqswelling}

When a gel with specific topology, chain length, and interaction parameters is to be studied, the equilibrium swelling, and hence the box size, is not known a priori.
Therefore, the first step in the analysis of the system is to determine this property in the field-free case.
In principle, it is possible to do this by simulating in the $NPT$-ensemble using a barostat. However, this was found not to be efficient\cite{mann05d}.
Instead, a set of simulations in the $NVT$-ensemble is performed at different volumes. By measuring the stress tensor and averaging over its diagonal elements, the stress-strain relation for the gel is obtained:
\begin{equation}
P(\epsilon) =\frac13 \sum_i S_{ii},
\end{equation}
where $P(\epsilon)$ denotes the isotropic pressure at a certain strain $\epsilon$, and $S_{ij}$ denote the elements of the stress tensor.
From this curve, the equilibrium strain, at which the stress is zero, can be estimated by linear interpolation.
Taking the box length of the initial system to be unity, with completely stretched polymers, the following equilibrium swelling was obtained:
for the diamond cubic lattice, values of $0.2790$ and $0.2546$ are found for chains with 60 and 80 beads, respectively.
The values for the simple cubic structure are $0.3153$ and $0.2742$.
It is convenient, to express the size of the gels in terms of this equilibrium swelling. This will be done throughout the rest of the paper.

\begin{figure}
\includegraphics[width=\stdFigWidth]{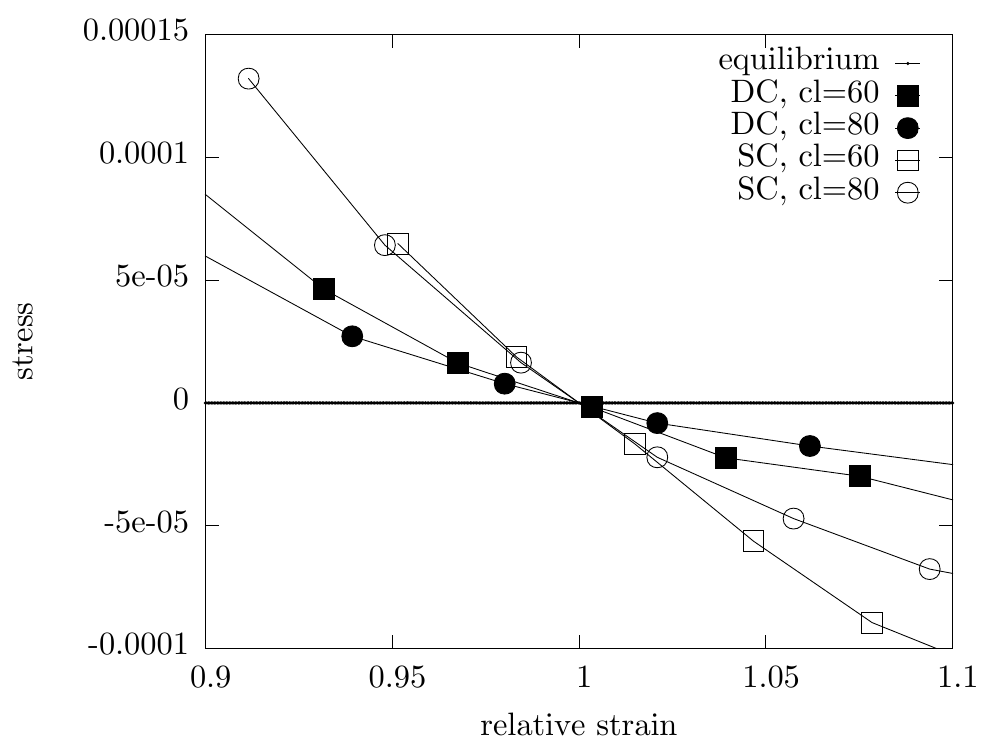}
\caption{\label{fig:fan-3d-stress-strain-rescaled}
Stress-strain relations for both, the diamond cubic (filled symbols) and the simple cubic (open symbols) geometry.
The $x$-axis is rescaled such that unity represents the equilibrium volume for the given geometry and chain length.
From the respective slopes of the stress-strain relations at the point of equilibrium swelling, the elasticity of the samples can be determined. It can be seen that for both chain lengths shown, the samples in the diamond cubic geometry are softer than those in the simple cubic geometry. Additionally, longer chains make the gel more deformable.
}
\end{figure}
In Fig.\,\ref{fig:fan-3d-stress-strain-rescaled}, the stress-strain curves for both, DC and SC geometries, are shown. 
From the slope of the curves at the equilibrium strain, the stiffness of the gel can be deduced. The higher the slope, the stiffer the material is. 
It can be seen that the gel in the SC geometry is more stiff than in the DC geometry. Also, a decrease of the chain length results in a stiffer system. Both these observations can be understood by noting the higher particle density for shorter chain lengths and for a larger number of chains.
A quantitative study of the elastic constants will follow in Sec.\,\ref{sec:el-const}.

\section{Anisotropic deformation}
\label{sec:aniso-def}

\begin{figure}
\includegraphics[width=\stdFigWidth]{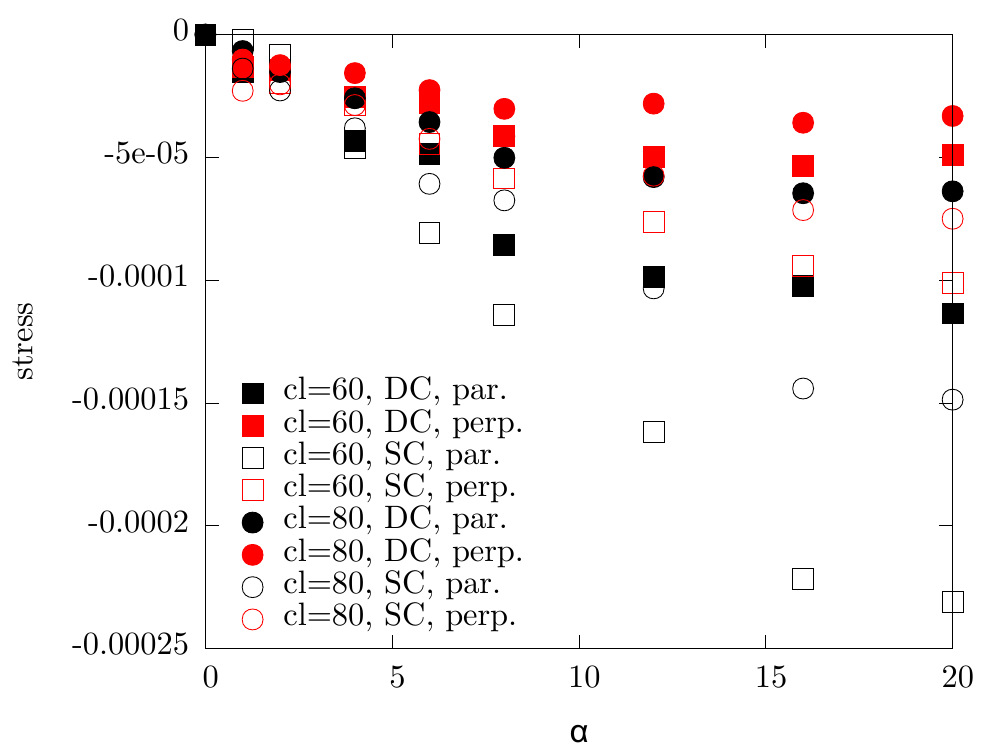}
\caption{\label{fig:stress-vs-field}
Stress measured parallel and perpendicular to the external field versus the strength of the external field. Data is shown for the diamond cubic (DC, full symbols) and simple cubic (SC, open symbols) geometry and for chain lengths (cl) of 60 and 80 beads per chain.
It can be seen that in both geometries, the absolute value of the stress parallel to the field is larger than the value measured in the perpendicular direction.
A much larger negative stress is observed in the simple cubic geometry than in the diamond cubic one.
}
\end{figure}

In this section, the deformation of the gel under the influence of an external field will be discussed. In particular, we will have a closer look at the anisotropic deformation described in Ref.\,\cite{weeber15a}.
As the first step, we measure the tendency of the gel to deform in the directions parallel and perpendicular to the field, due to the wrapping of the polymer chains around the magnetic particles (Fig.\,\ref{fig:sketch}).
In order to separate this wrapping effect from the elastic properties, we first measure the direction-dependent stress due to an external field in the non-deformed gel.
Hence, the gel is simulated in the $NVT$-ensemble at the equilibrium volume for the field free case as it was determined in the previous section.
Then, a field is applied and the stress in the directions parallel and perpendicular to the field is measured.
The results for chain lengths of 60 and 80 beads, and for both, the DC and SC geometry, are shown in Fig.\,\ref{fig:stress-vs-field}.
Throughout this paper, magnetic fields are expressed in terms of the Langevin parameter
\begin{equation}
\label{eqn:alpha}
\alpha =\frac{\mu_0 m H}{k_B T},
\end{equation}
where $m$ denotes the particles' moment, $\mu_0$ the vacuum permittivity and $H$ the external field. This parameter measures the ratio between the Zeeman energy and the thermal energy $k_B T$.
The measured stresses for all applied external magnetic fields larger than zero are negative, indicating that the gels would contract if the constraint of the constant volume would be removed.
The absolute value of the stress in the direction parallel to the field is significantly larger than the one measured in the perpendicular direction.
It is also worth noting that the stress is significantly stronger in the SC geometry, where six chains are connected to a node as opposed to four in the DC geometry.
This implies that two contradicting forces are at play. On the one hand, in the simple cubic structure, there are more chains attached to a node. This leads to a stronger wrapping effect and a stronger tendency of the system to deform in the field. On the other hand, from the slope at the equilibrium swelling in Fig.\,\ref{fig:stress-vs-field}, it can be observed that the gel based on the simple cubic structure is stiffer than the one based on the diamond structure.
From measurements of the deformation of the gel, it turns out that for this particular model, the elasticity dominates the deformation behaviour.

\begin{figure*}
\includegraphics[width=\stdFigWidth]{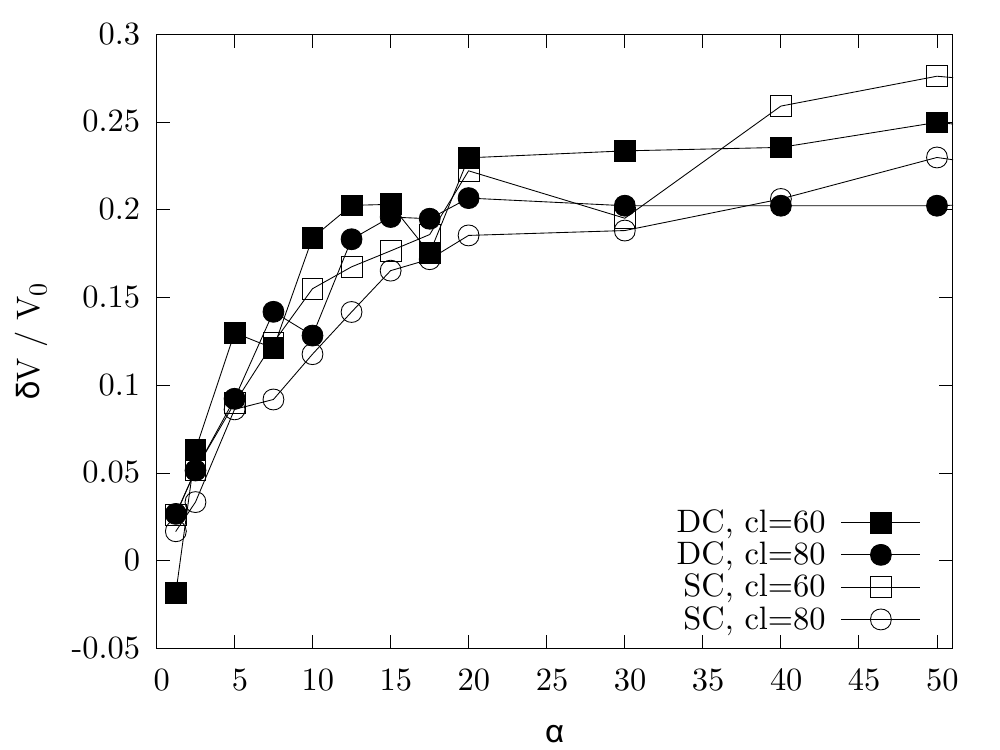}
\includegraphics[width=\stdFigWidth]{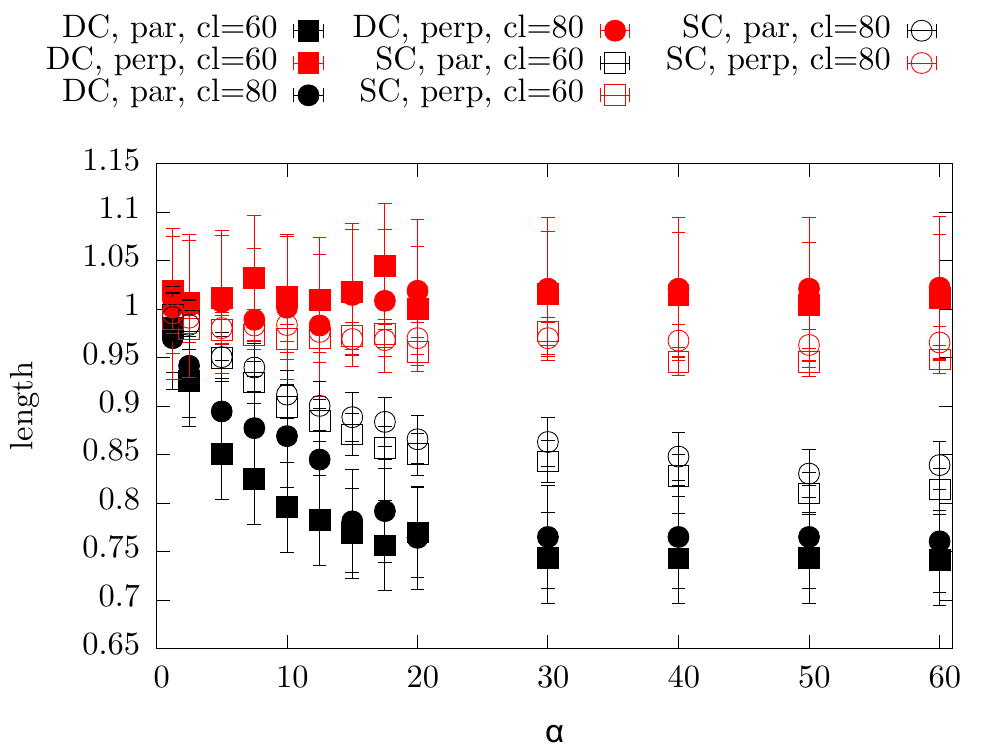}
\caption{\label{fig:fan-3d-shape}
Left: relative loss of volume of a gel versus the strength of the external field.
Right: length of the gel in the directions parallel and perpendicular to an externally applied magnetic field versus the strength of that field.
Full symbols denote the diamond cubic (DC) geometry, whereas open symbols are used for the simple cubic (SC) one. The chain length (cl) was 60 (squares) and 80 (circles), respectively.
It can be seen that the overall shrinkage is comparable in magnitude for all systems considered within the simulation error, which, however, is large.
While all samples shrink significantly more in the direction parallel to the field, it is noteworthy that the gels in the simple cubic geometry shrink less in the direction parallel to the field, but more in the perpendicular direction than those in the diamond cubic geometry. This results in a similar change of volume in all systems, although the ``shape`` of the deformation depends strongly on the network geometry.
}
\end{figure*}
To find the equilibrium shape of the gel for a given external magnetic field, 
simulations in the $NVT$-ensemble are executed iteratively. In each simulation, the stress is measured in the directions parallel and perpendicular to the external field. Then, the box size for the next simulation is altered according to these stress measurements.
This procedure is repeated until the remaining stress approaches zero within the chosen error tolerance.
The details of the procedure are as follows:
\begin{itemize}
\item A shape of a simulation consists of the lengths of the simulation box parallel and perpendicular to the external field.
\item For each shape simulated, the stress measurement is averaged over 24 simulations to reduce the measurement error.
\item Initially, two sets of simulations are performed for a given external magnetic field. The gel samples in them have manually selected shapes, of which one is above and the other is below the expected equilibrium shape for the given field. For these shapes, the corresponding stresses in the direction parallel and perpendicular to the field are measured.
\item In all further iterations, a single new shape consisting of a box length parallel and perpendicular to the field is generated.
For the respective directions parallel and perpendicular to the field,
a new box length is generated from the two ones examined in previous iterations which yielded the lowest corresponding absolute stress. 
From these box lengths and corresponding stresses, a linear stress-strain relation is constructed, which is used to extrapolate to the point where the stress would be zero according to this relation. 
Due to large statistical errors in the stress measurements, it is necessary to restrict the change in box length to approximately two percent per iteration, to keep the procedure stable.
For this new shape, again, the stress is measured in the directions parallel and perpendicular to the external magnetic field. 
\item The iteration is terminated, if the stress in both, the direction parallel and perpendicular to the field, is below $10^{-5}$.
\end{itemize}
The iterative procedure provides satisfactory results but requires a large amount of computational resources and occasional manual interventions. 
In a future study, more sophisticated procedures, like genetic optimizations might be considered.
As an alternative to the iterative procedure, it is also possible to simulate a set of shapes around the expected equilibrium shape, obtain the corresponding stresses, and fit a linear stress-strain model to the data. This is done in conjunction with the measurement of elastic constants in Sec.\,\ref{sec:el-const}.

To study the deformation of the gel in an external magnetic field, the equilibrium shapes were determined for fields up to $\alpha=60$ in terms of the Langevin parameter.
Results for chain lengths of 60 and 80 beads are shown on the right hand side of Fig.\,\ref{fig:fan-3d-shape} for gels based on both, the diamond cubic and simple cubic network structure. 
It can be seen that, as expected from the anisotropic stress measurements at constant volume (Fig.\,,\ref{fig:stress-vs-field}) and as explained above,
the gel deforms significantly more in the direction parallel to the external field than in the perpendicular direction.
The deformation decreases when the chain length is increased.
This can be explained by noting that, as the node particles align to the external field, the chains attached to the node are effectively shortened.
The entropic effect of this shortening is more significant for a short chain than for a long one. While a long chain can slightly uncoil to accommodate for the motion of its ends, a short chain will be forced into a rather straight configuration, which is entropically very unfavorable.

The influence of the network geometry (diamond cubic vs. simple cubic) on the amount of shrinkage is determined by three factors.
First, more chains attached to a node lead to a higher stress, when a field is applied, as more chains get stretched by the node particles' rotation.
Second, more chains lead to a higher density of particles in the system, and therefore also a higher stiffness. 
Finally, a contraction in one direction in most materials leads to an expansion in the perpendicular directions. The strength of this expansion in the direction perpendicular to the stress is measured by the Poisson ratio (Eqn.\,\ref{eqn:poissondef}),  which also depends on the network geometry.

As can be seen from the plot of the gel's equilibrium shape versus the field (right hand side of Fig.\,\ref{fig:fan-3d-shape}), 
the amount of shrinkage in the direction parallel to the field is larger for the diamond cubic geometry than for the simple cubic one.
In the direction perpendicular to the field, gels based on the diamond cubic geometry expand slightly, while they contract when based on the simple cubic geometry.
The relative shrinkage, i.e., the loss of volume due to an applied field, divided by the volume in the field free case is given by
\begin{equation}
\frac{\delta V(\alpha)}{V_0} =1 -\frac{l_{\rm{par}}(\alpha) l_{\rm{perp}}(\alpha)^2}{l_0^3},
\end{equation}
where $l_{\rm{par}}(\alpha)$ and $l_{\rm{perp}}(\alpha)$ denote the length of the gel parallel and perpendicular to an external magnetic field with a Langevin parameter of $\alpha$, and $l_0$ denotes the length of the gel in the field free case, which is equal in all Cartesian directions.
The shrinkage for gels based on the diamond cubic and simple cubic geometries and chain lengths of 60 and 80 beads is shown on the left hand side of Fig.\,\ref{fig:fan-3d-shape}.
It can be seen that the shrinkage is of similar magnitude for all systems considered, even though the actual shape of the gel (Fig.\,\ref{fig:fan-3d-shape}) depends strongly on the network geometry.
The similarity of the shrinkage stems from the fact that, while the simple cubic gels shrink less in the direction parallel to the field than the diamond cubic ones, they shrink more in the perpendicular direction.

\section{Elastic constants and Poisson ratio}
\label{sec:el-const}

To gain an understanding of the deformation of magnetic gels in external fields, it is not only necessary to study the underlying mechanism. Additionally, the elastic properties of the material need to be considered, as they determine, to what degree the material will deform under a given stress.
Additionally, in an external magnetic field, not only the shape of a magnetic gel can change, but also its elastic properties.
Hence, in this section, the change in these elastic constants will be examined for our two gel models.
The resulting stress can be assumed to be linear in the strains, if the strains are sufficiently small.
In this case, the elastic constants describe what kind of stress is observed in a system, when a certain strain or shear is applied.
When only the linear strain but no shear is considered, the elastic constants can be written down in  matrix form as follows
\begin{equation}
C =
\left(
\begin{array}{ccc}
c_{xx} & c_{xy} & c_{xz} \\
c_{yx} & c_{yy} & c_{yz} \\
c_{zx} & c_{zy} & c_{zz}
\end{array}
\right),
\end{equation}
where the first index denotes the direction of the stress response and the second index denotes the direction of a strain.
Then, for a given strain in $x$, $y$ and $z$-direction
\begin{equation}
\epsilon = 
\left(
\begin{array}{c}
\epsilon_{xx} \\
\epsilon_{yy} \\
\epsilon_{zz}
\end{array}
\right).
\end{equation}
the resulting stress is
\begin{equation}
\sigma =C \epsilon.
\end{equation}
Hence, to describe the elastic properties of the gel in the linear response regime, the elasticity matrix $C$ has to be determined.

Depending on the symmetry of the system, some of the entries of the elastic matrix $C$ are identical.
For an isotropic material, there are only two free parameters, namely the diagonal and the off-diagonal elements of $C$.
The diagonal elements describe the stress in the direction parallel to the strain, whereas the off-diagonal elements describe the strain in the perpendicular direction. We have
\begin{equation}
C =
\left(
\begin{array}{ccc}
a &b &b\\
b &a &b\\
b &b &a
\end{array}
\right).
\end{equation}
This situation applies to the gel, when no external field is applied.

When, on the other hand, a field is applied,
a distinction needs to be made between the direction parallel to the field and the two Cartesian directions perpendicular to it. 
When the external magnetic field is parallel to the $x$-direction,  the resulting elastic matrix has the from
\begin{equation}
\label{eqn:el-const-model-field}
C_{\rm{field}} =
\left(
\begin{array}{ccc}
a & b & b\\
c & d & e\\
c & e & d
\end{array}
\right).
\end{equation}
As it can be seen, there are five independent elastic constants.
The constant $a$ describes the stress parallel to the field for a strain parallel to the field, $b$ describes the stress parallel to the field for a strain perpendicular to the field, and $c$ describes the stress perpendicular to the field for a strain parallel to the field.
The symbol $d$ describes the stress parallel to a strain which is applied perpendicular to the field.
Finally, $e$ describes the stress in a direction perpendicular to the field, when the strain is applied along the second Cartesian direction perpendicular to the field.

In this paper, the elastic constants are obtained by numerically calculating the derivatives of the stress tensor by running simulations at different strains close to the equilibrium strain, measuring the stress, and fitting a linear model to the resulting data.
The technical details of this procedure can be found in the supplementary information.

The elastic constants are obtained for gels based on both, the simple cubic and the diamond cubic network geometries with a chain length of 60 beads. 
Measurements are performed for the field-free case as well as for an external field of $\alpha=20$. Even when no field is applied, no additional constraints are applied to the elasticity matrix. 

\begin{figure}
\input{compare-geometry.inc}
\caption{\label{fig:fan-3d-el-const-geometry}
Comparison of the elastic constants $a$ through $e$ as defined by Eqn.\,\ref{eqn:el-const-model-field} for gels constructed in the diamond cubic and simple cubic geometries for the field-free case ($\alpha = 0$).
It can be seen, that the absolute value of the constants for the simple cubic system are always larger than those for the diamond cubic one. This indicates a more rigid gel.
}
\end{figure}
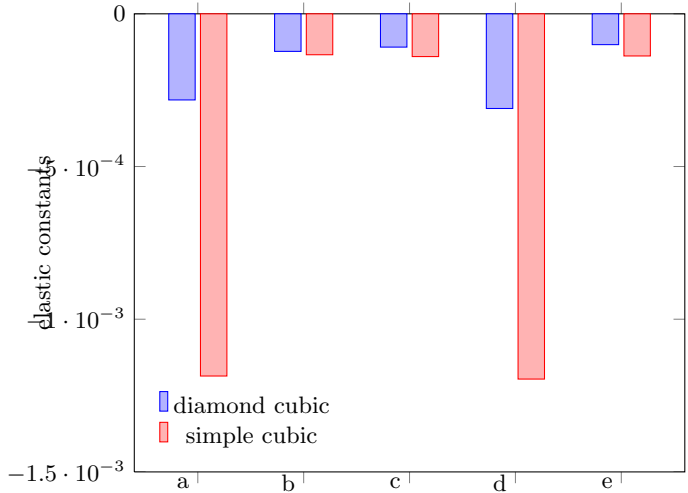

In Fig.\,\ref{fig:fan-3d-el-const-geometry}, the elastic constants are compared for gels in the simple cubic and diamond cubic geometries.
It can be seen that gels in the simple cubic geometry are significantly stiffer than those in the diamond cubic one.
In particular, the elastic constants which describe a stress occurring parallel to a strain ($a$ and $d$ in Eqn.\,\ref{eqn:el-const-model-field}) are approximately four times larger. The off-diagonal elements of the elasticity matrix ($b$, $c$, and $e$) are only higher by 10 to 25\% in the simple cubic geometry.

\begin{figure}
\input{compare-field-dc.inc}
\newline
Diamond cubic\\

\input{compare-field-sc.inc}
\newline
Simple cubic
\caption{
\label{fig:fan-3d-el-const-field}
Comparison of the elastic constants $a$ through $e$ (Eqn.\,\ref{eqn:el-const-model-field}) of our model gel for the cases of no external field and an external field of $\alpha=20$. Results for the diamond cubic geometry are shown at the top, while results for the simple cubic geometry are shown at the bottom.
}
\end{figure}
In Fig.\,\ref{fig:fan-3d-el-const-field}, the elasticity of a gel in an external field ($\alpha=20$) is compared to that of a gel with no applied field. Data is presented for the DC and SC geometries, respectively.
We observe that all ratios are larger than unity. Hence, the material is more stiff, when an external magnetic field is applied. This can be understood by noting that the gel shrinks in a field, and therefore the density of particles in the system is higher.
It is notable that in the simple cubic geometry, the off-diagonal elements ($b$, $c$ and $e$) are increased more strongly than the diagonal elements ($a$ and $d$), when an external field is applied.

From the measurement of the elasticity matrix, it is also possible to calculate the Poisson ratio for the investigated systems.
For a prescribed strain in one direction, these ratios measure the resulting strain in the perpendicular directions.
In the case of a system with one spacial direction, namely the field direction, there are three different Poisson ratios:
a prescribed strain in the field direction and a resulting strain in a non-field direction, a prescribed strain in a non-field direction and a resulting strain in the field direction, and a prescribed strain in a non-field direction and a resulting strain in the second non-field direction.
The subscript $f$ and $n$ will be used to denote field and non-field directions, respectively.
The first letter denotes the direction of the prescribed strain, whereas the second letter denotes the direction of the response.
The Poisson ratio is positive, when a prescribed expansion leads to a contraction in the response direction.
The field is assumed to be applied in $x$-direction.
To obtain the Poisson ratio $p_{fn}$ for a prescribed strain in field direction and a resulting strain in a non-field direction, we apply the strain
\begin{equation}
\label{eqn:poissondef}
\vec{\epsilon} =\left(\begin{array}{c}
1 \\ -p_{fn} \\ -p_{fn} 
\end{array}
\right)
\end{equation}
to the system. Note that for a prescribed strain of unity, the resulting contraction in the perpendicular directions is the Poisson ratio.
Now, to find the Poisson ratio, we require that for the given $\vec{\epsilon}$, the stress in the response directions has to be zero:
\begin{equation}
C\vec{\epsilon} =\left(\begin{array}{c}
x\\0\\0
\end{array}\right),
\end{equation}
where $C$ is the elasticity matrix. 
The stress occurring in the direction of the prescribed strain is irrelevant, so $x$ can take an arbitrary value. In other words, only the second and third row of the equation have to be solved.
Using Eqn.\,\ref{eqn:el-const-model-field} for the elasticity matrix $C$, we find
\begin{equation}
p_{fn} =\frac{c}{d+e}.
\end{equation}
The remaining two Poisson ratios, $p_{nf}$ and $p_{nn}$, can be obtained similarly. They are
\begin{equation}
p_{nf} =\frac{b}{a+b}\\
p_{nn} =\frac{e}{d+e}.
\end{equation}

\begin{figure}
\input{poisson.inc}
\caption{\label{fig:poisson}
Poisson ratios $p_{fn}$, $p_{nf}$ and $p_{nn}$ for gel samples based on the diamond cubic (DC) and simple cubic (SC) geometries for external magnetic fields of $\alpha=0$ and $\alpha=20$.
In the field-free case, all Poisson ratios for a given geometry are expected to be the same. The visible deviations are due to statistical measurement errors. 
It can be see that the direction of the prescribed strain as well as the direction of the measured response do not significantly influence the Poisson ratio. The Poisson ratio does, however, strongly depend on the network geometry. In the diamond cubic case, a much higher value is observed.
}
\end{figure}
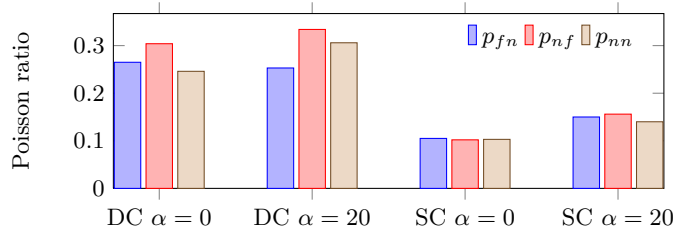

Based on these equations, the Poisson ratios for the gels can be calculated using the elastic constants shown in Fig.\,\ref{fig:fan-3d-el-const-field}.
The resulting ratios are shown in Fig.\ref{fig:poisson}.
Values are shown for networks based on the diamond cubic and simple cubic geometry, and for both, the field-free case and an external field of $\alpha=20$.
In the field-free case, the system is isotropic, and thus the three Poisson ratios $p_{fn}$, $p_{nf}$ and $p_{nn}$ are expected to be equal. In the figure, however, a deviation of about 20\% is visible due to statistical errors.
It can be seen that the choice of directions for the prescribed strain and the observed response (field or non-field direction) does not have a strong influence on the Poisson ratio.
The choice of network geometry -- diamond cubic or simple cubic --, on the other hand, has a strong influence on the Poisson ratio: a much higher value is observed for the diamond cubic geometry. The Poisson ratios also increase, when a strong magnetic field is applied.

In summary, a procedure has been described to obtain the elastic constants by fitting a linear elastic model to a set of stress-strain measurements. The elastic constants are found to be influenced mostly by the choice of network topology. In the diamond cubic case, the gel is much softer but has much larger Poisson ratios compared to the simple cubic case.
The elastic constants turn out not to be strongly anisotropic, even when an external field is applied.
The differences in Poisson ratio help to explain the differences in the deformation behaviour found for different network geometries (Fig.\,\ref{fig:fan-3d-shape}).
In particular, the low shrinkage in the direction perpendicular to the field for the gel based on the diamond cubic geometry is related to the high Poisson ratio for this system:
Due to the shrinkage in the direction parallel to the field, an expansion occurs in the perpendicular direction. Due to the high value of the Poisson ratio for this geometry, this compensates for the shrinkage which would otherwise occur due to the rolling-up of polymer chains around the magnetic particles.

\section{Magnetic response}
\label{sec:mag}

\begin{figure}
\begin{center}
\includegraphics[width=\stdFigWidth]{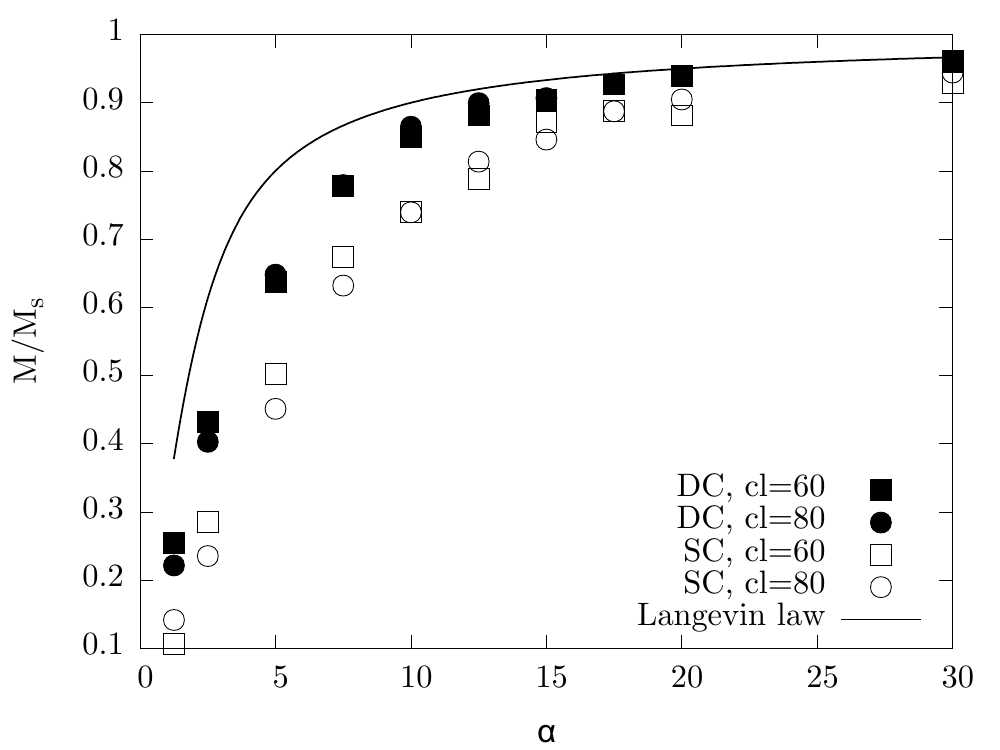}
\end{center}
\caption{\label{fig:fan-3d-mag-curves}
Magnetization curve for gels of the diamond cubic and simple cubic geometries for chain lengths of 60 and 80 beads, respectively. The solid line shows the magnetization curve expected for non-interacting dipoles in three dimensions (Langevin curve).
It can be seen that the magnetization for all gels considered is below that of non-interacting dipoles.
As soon as the field aligns the magnetic node particles of the polymer network, a stress is created on the polymer chains.
Due to this stress, the polymer chains exert a torque on the magnetic particle which acts against the alignment to the external field.
The torque counteracting the magnetization is larger, if the chains are shorter, and if there are more chains. Thus, a lower magnetization is observed for the case of 60 beads as well as the case of the simple cubic network geometry.
}
\end{figure}

In the model considered here, through the binding of the polymer chains to specific spots on the surface of the magnetic nanoparticles cross-linking the network, a coupling is created between the magnetic particles' orientation and the polymer matrix.
This coupling is the basis of the deformation mechanism for this kind of magnetic gels.
By means of the gel's magnetic response to an external magnetic field, further insights into the details of the deformation mechanism can be gained. 
In Fig.\,\ref{fig:fan-3d-mag-curves}, the magnetization curve
$M(\alpha)$ is shown for gels based on the diamond cubic and simple cubic geometries with chain lengths of 60 and 80 beads.
The curve describes the component parallel to the magnetic field of the sum of all magnetic moments in the system
for a given field expressed in terms of the Langevin parameter (Eqn.\,\ref{eqn:alpha}).
Also shown is the Langevin law, which is the result expected for non-interacting dipoles, given by
\begin{equation}
M_l(\alpha) = \frac{\cosh{\alpha}}{\sinh{\alpha}} - \frac{1}{\alpha}.
\end{equation}
From the plot, it can be seen that the magnetization of the gels is always lower than the corresponding Langevin magnetization. This is due to the fact that the polymer chains get strained, when the magnetic particles rotate to align their dipole moments to the external field.
The strain on the polymer chains causes a stress on them, which in turn creates a torque on the magnetic particles, which counteracts the magnetic field.
This leads to a reduction in magnetization.
The loss of entropy per chain, due to its stretching, increases for shorter chains. 
Thus, a larger stress occurs and the magnetization is lower for the sample with shorter chains. Also, the torque on the magnetic particles is larger, when more chains are attached to it. Hence, the magnetization is also lower for the gel based on the simple cubic geometry, in  which six chains are connected to each node particle, compared to four chains attached in the diamond cubic geometry.

\section{Summary}
\label{sec:gels-summary}

In this paper, we presented a detailed study of the deformation mechanism in  a magnetic gel which is cross-linked by magnetic nanoparticles.
The deformation mechanism described arises from a direct coupling of the orientational degree of freedom of the magnetic moments to the polymer chains.
When an external field is applied, the magnetic nanoparticles rotate and exert a stress on the polymer chains attached to them. This, in turn leads to a contraction of the matrix.
The model, which we study by means of coarse-grained molecular dynamics, is inspired by an experimental system discussed in Ref.\,\cite{messing11a}.
In the three dimensional case, which is the focus of this report, an anisotropic deformation is observed. 
While there is a strong shrinkage in the direction parallel to the field, the shrinkage in the perpendicular directions is either small or not present at all depending on the network topology.

A comparison was made between two network topologies, one with four chains and another one with six chains connected to a node. In these topologies, the nodes are arranged in a diamond cubic and simple cubic geometry, respectively.
An increase of the number of chains connected to a node has two effects. On the one hand, it increases the stress on the gel caused by an alignment of the magnetic particles to the external field. This, on its own, should lead to an increased shrinkage of the gel. However, at the same time the higher number of chains results in a higher particle density in the model gel, which should lead to a lower shrinkage. For the systems studied here, there is a stronger shrinkage, when four chains are attached to a node.
As this result is presumably highly dependent of the details of the model, it is conceivable that the trade-off between the two mentioned trends can lead to a different behaviour for a system with, for instance, significantly lower overall density.
Elastic properties were studied by fitting a linear elasticity model to a set of strain measurement. We found a significantly higher stiffness for the more strongly cross-linked gels in the simple cubic structure, as well as an increase of the stiffness when a field is applied.

The magnetic response of the model gel which is cross-linked by magnetic particles is strongly influenced by the coupling between the orientation of the magnetic nanoparticles and the polymers. As the alignment of the magnetic particles to an external field exerts a stress on the polymers, there is an additional energy penalty for this alignment. In consequence, the magnetic response of the gel is below that of non-interacting magnetic particles.

In summary, a detailed study of the deformation, elasticity and magnetic response of a particle-cross-linked ferrogel was performed using simulations.
Based on the results, we have shown that the overall deformational response is determined by an interplay between the gel's degree of cross-linking and its elasticity.
In the future, we plan to extend the model by introducing randomness into the chain lengths and the connectivities.
Additionally, in experimental systems, different deformation mechanisms might occur at the same time. This would be the case for gels, which are cross-linked by magnetic particles with a very high magnetic moment and at a high density.

\section*{Acknowledgement}
RW and CH are grateful for financial support from the DFG through the
SPP 1681. In addition, we acknowledge funding through the cluster of
excellence EXC 310, SimTech, and access to the computer facilities of
the HLRS and BW-Unicluster.  SK was supported by Austrian Science Fund
(FWF): START-Project Y 627-N27 and RFBR grant mol-a-ved 12-02-33106.

\bibliographystyle{unsrt}
\bibliography{ferrogels}

\newpage

\section*{Supplementary information}

\subsection*{Technical details of the determination of the elastic constants}

In the paper, the elastic constants are obtained by numerically calculating the derivatives of the stress tensor by running simulations at different strains close to the equilibrium strain, measuring the stress, and fitting a linear model to the resulting data.

To obtain the data, the stress is measured in two series, namely
\begin{itemize}
\item a set of simulations in which one strain value is applied to the Cartesian axis parallel to the field, and another strain is applied to both directions perpendicular to the field
\item a set of simulations, in which the Cartesian axis parallel to the field and one of the axes perpendicular to the field are unstrained, but the other axis perpendicular to the field is strained.
\end{itemize}
The relative strains applied are $\pm 3\%$ and $\pm 5\%$ as well as $0$,
resulting in 30 strain configurations.
For each strain, eight simulations were performed for averaging. 
During the simulations, stress measurements are taken every four time steps.
There are approximately 200\,000 samples per simulation in the simple cubic geometry and 400~000 samples for the diamond cubic geometry.
The remaining simulation parameters have been described in Sec.\,\ref{sec:sim}.
The size of the unstrained system for these simulations is taken from the iterative procedure described in Sec.\,\ref{sec:aniso-def}.
The equilibrium size (Fig.\,\ref{fig:fan-3d-shape})
is characterized by two parameters, $l_{\rm{par}}$ and $l_{\rm{per}}$, which denote the length of the simulation box in the directions parallel and perpendicular to the applied magnetic field.
A length of $1$ refers to a simulation box, in which the polymer chains between the nodes of the network are in a straight line configuration and the spacing between neighboring beads equals the Lennard-Jones $\sigma$.

The error bar on these measurements of the equilibrium size for a given field is significant.
However, these values need to be known to calculate relative strains in terms of the equilibrium size.
To circumvent this problem, the equilibrium size is allowed to vary during the fit.
This way, an independent estimate for the equilibrium swelling can be determined from the measurements of the stress-strain relation.

In total, there are seven fit parameters
\begin{equation}
p =(l_{\rm{par}},l_{\rm{per}},a,b,c,d,e),
\end{equation}
where $l_{\rm{par}}$ and $l_{\rm{per}}$ are the equilibrium size of the system, as described above, and $a$ through $e$ are the five independent entries of the elastic matrix as in Eqn.\,\ref{eqn:el-const-model-field}.
The target function of the fit is given by
\begin{equation}
\label{eqn:el-const-target-function}
F(p) =\frac1N \sum_k | \vec{\sigma}^{\rm{m}}_k -\vec{\sigma}^{rm{c}}_k|^2,
\end{equation}
where the sum is over all $N$ strains for which a stress has been measured.
$\vec{\sigma}^{\rm{m}}_k$ denotes the stress measured in the simulation
and 
$\vec{\sigma}^{\rm{c}}_k$ denotes the stress calculated using the current estimate of the fit parameters $p$.
That one is obtained by constructing the elasticity matrix $C$ from the current estimates of the fit parameters $a$ through $e$.
Then the relative strain $\vec{\epsilon}_k$ for a given simulation $k$ is calculated from the ratio of its actual simulation box and the current estimates of the equilibrium box size $l_{\rm{par}}$ and $l_{\rm{perp}}$.
The calculated stress for this simulation is then
\begin{equation}
\vec{\sigma}_k(p) =C(p) \vec{\epsilon}_k(p).
\end{equation}

\begin{figure*}
\begin{center}
\includegraphics[width=5.8cm]{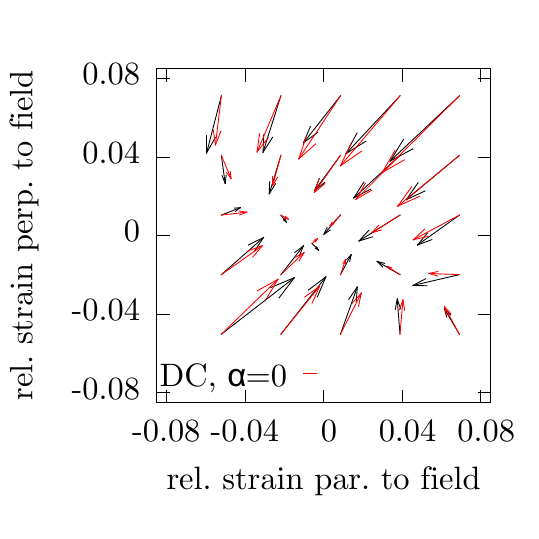}
\includegraphics[width=5.8cm]{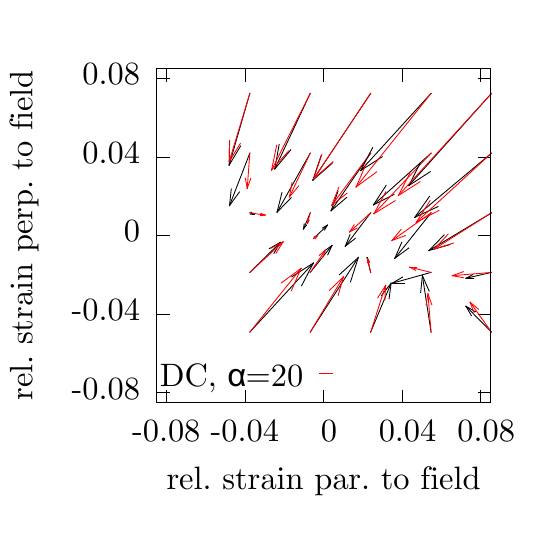}
\\
\vspace{0.3cm}
\includegraphics[width=5.8cm]{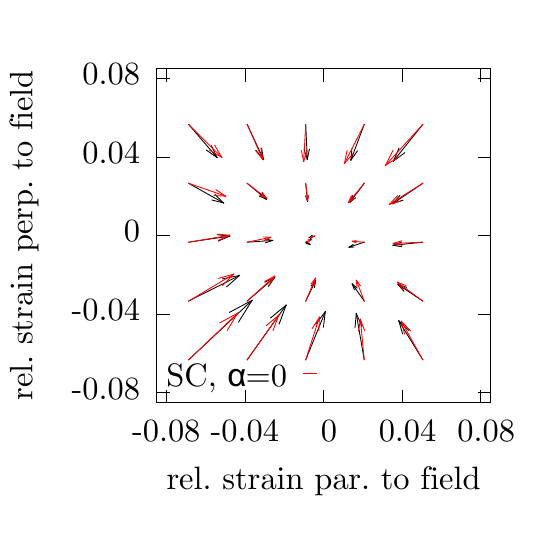}
\includegraphics[width=5.8cm]{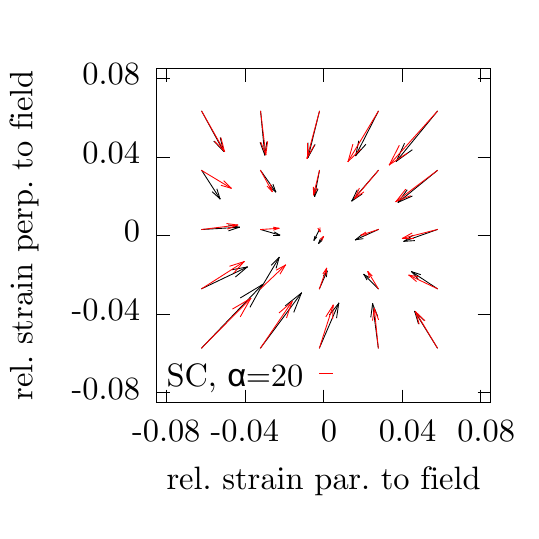}
\end{center}
\caption{\label{fig:fan-3d-stress-fit}
Plot of the measured stress (black arrows) and stresses calculated based on the fitted elasticity matrix (red arrows).
Results are shown for configurations which have a strain applied to the axis parallel to the magnetic field and an other strain applied to both axes perpendicular to the field.
}
\end{figure*}

To perform the actual fit,
the target function, Eqn.\,\ref{eqn:el-const-target-function} is minimized using the l\_bfgs\_b constraint minimization of the scientific python library.
The entries of the elastic matrix are allowed to assume a value between $-1$ and $1$. The equilibrium size is allowed to deviate by $0.02$ from the estimate obtained from the iterative procedure (Fig.\,\ref{fig:fan-3d-shape}).
In Fig.\,\ref{fig:fan-3d-stress-fit},
the measured and calculated stresses are shown for the simulations in which the strain on the two axes perpendicular to the field is equal.
It can be seen that there is qualitative agreement between measured and calculated stresses, but that there are some outliers.
There are two causes for this, namely a large statistical error in the measured stress, and fitting errors.

The quality of the fit can then be evaluated by verifying that those components of the elasticity matrix which should have identical values due to the system symmetry actually are similar.

To asses the quality of the fits, the following checks are preformed
\begin{itemize}
\item The equilibrium sizes as determined by the fit agree within approximately two percent with those obtained from the iterative procedure.
\item The entries of the elastic matrix are negative. When applying a positive strain, i.e., when expanding the system, the stress is negative.
\item When there is no magnetic field applied
\begin{itemize}
\item The equilibrium size is the same in along all Cartesian axes ($l_{\rm{par}} \approx l_{\rm{perp}}$).
\item Because of the isotropy in the field free case, only two values in the elastic matrix can be chosen freely, namely the value of the diagonal elements and the value of the off-diagonal elements. 
Even when we fit the full linear model, these constants need to assume similar values, if the fitting result is to be physically meaningful.
I.e., in Eqn,\,\ref{eqn:el-const-model-field}, $a\approx d$ as well as $b\approx c \approx e$.
This is the case up to an accuracy of approximately ten percent for the diamond cubic geometry and approximately three percent for the simple cubic geometry.
\end{itemize}
\end{itemize}

The resulting elastic constants are summarized in table \ref{tbl:el-const-summary}. they are also shown in plots \ref{fig:fan-3d-el-const-geometry} through \ref{fig:poisson} in the paper.

\begin{table}
\begin{tabular}{|lll|ll|p{1.1cm}p{1.1cm}p{1.1cm}p{1.1cm}p{1.1cm}|}
\hline
geo.&$l_c$&$\alpha$&$l_{\rm{par}}$&$l_{\rm{per}}$&$c_{xx}$\newline($a$)&$c_{xy}$$=$$c_{xz}$\newline($b$)&$c_{yx}$$=$$c_{zx}$\newline($c$)&$c_{yy}$$=$$c_{zz}$\newline($d$)&$c_{yz}$$=$$c_{zy}$\newline($e$)\\
\hline
DC&60&0&0.277&0.276&-0.282&-0.123&-0.109&-0.310&-0.101\\
\hline
DC&60&20&0.215&0.276&-0.408&-0.205&-0.190&-0.522&-0.230\\
\hline
SC&60&0&0.313&0.311&-1.186&-0.134&-0.140&-1.196&-0.131\\
\hline
SC&60&20&0.269&0.301&-1.623&-0.300&-0.308&-1.763&-0.287\\
\hline
\end{tabular}
\caption{
\label{tbl:el-const-summary}
Elastic constants for a gel sample in the diamond cubic (DC) and simple cubic (SC) network geometries for a chain length of $l_c=60$ and external magnetic fields $\alpha=0$ and $\alpha=20$.
$l_{\rm{par}}$ and $l_{\rm{per}}$ denote the equilibrium size of the gel in the directions parallel and perpendicular to the applied field.
The elastic constants $c_{ij}$ denote the stress observed in the $i$-direction when the system is strained in the $j$-direction.
Due to the fact that the two directions perpendicular to the field are indistinguishable, some of the constants are equal and are shown only once.
The letters in brackets refer to the symbols in Eqn.\,\ref{eqn:el-const-model-field}.
}
\end{table}

\end{document}

%% file: compare-geometry.inc
\begin{tikzpicture}
\begin{axis}[
    ybar,
    enlargelimits=false,
    enlarge x limits=0.15,
    legend pos=south west,
    legend style={draw=none},
    ylabel={elastic constants},
    symbolic x coords={a,b,c,d,e},
    xtick=data,
    x tick label style={anchor=east},
    width=\stdFigWidth,
    ymin=-1.5E-3,
    ymax=0
    ]
\addplot+[ybar] plot coordinates {

(a,-0.000282) (b,-0.000123) (c,-0.000109) (d,-0.000310) (e,-0.0001009)
};
\addplot+[ybar] plot coordinates {

(a,-0.001186) (b,-0.000134) (c,-0.000140) (d,-0.001196) (e,-0.000138)
  };
\legend{diamond cubic, simple cubic}
\end{axis}
\end{tikzpicture}

%% file: compare-field-dc.inc
\begin{tikzpicture}
\begin{axis}[
    ybar,
    enlargelimits=false,
    enlarge x limits=0.15,
    legend pos=south west,
    legend style={draw=none},
    ylabel={elastic constants},
    symbolic x coords={a,b,c,d,e},
    xtick=data,
    x tick label style={anchor=east},
    width=\stdFigWidth,height=8cm,
    ymax=0
    ]
\addplot+[ybar] plot coordinates {

(a,-0.000282) (b,-0.000123) (c,-0.000109) (d,-0.000310) (e,-0.0001009)
};
\addplot+[ybar] plot coordinates {
(a,-0.000408) (b,-0.000205) (c,-0.000190) (d,-0.000522) (e,-0.000230)
  };
\legend{$\alpha=0$, $\alpha=20$}
\end{axis}
\end{tikzpicture}

%% file: compare-field-sc.inc
\begin{tikzpicture}
\begin{axis}[
    ybar,
    enlargelimits=false,
    enlarge x limits=0.15,
    legend pos=south west,
    legend style={draw=none},
    ylabel={elastic constants},
    symbolic x coords={a,b,c,d,e},
    xtick=data,
    x tick label style={anchor=east},
    width=\stdFigWidth,height=8cm,
    ymax=0
    ]
\addplot+[ybar] plot coordinates {
(a,-0.001186) (b,-0.000134) (c,-0.000140) (d,-0.001196) (e,-0.000138)
};
\addplot+[ybar] plot coordinates {
(a,-0.001623) (b,-0.000300) (c,-0.000308) (d,-0.001763) (e,-0.000287)
  };
\legend{$\alpha=0$, $\alpha=20$}
\end{axis}
\end{tikzpicture}

%% file: poisson.inc
\begin{tikzpicture}
\begin{axis}[
    ybar,
    legend style= { draw=none, 
      legend columns=-1},
    legend pos=north east,
    ylabel={Poisson ratio},
    xticklabels={DC $\alpha=0$, DC $\alpha=20$, SC $\alpha=0$, SC $\alpha=20$},
    xtick=data,
    ymin=0,
    width=\stdFigWidth,
    height=0.45\linewidth,
    ]
\addplot+[ybar] plot coordinates {
(1,0.265) (2,0.253) (3,0.105) (4,0.150)
};
\addplot+[ybar] plot coordinates {
(1,0.304) (2,0.334) (3,0.102) (4,0.156)
};

\addplot+[ybar] plot coordinates {
(1,0.246) (2,0.306) (3,0.103) (4,0.14)
};

\legend{$p_{fn}$, $p_{nf}$, $p_{nn}$},
\end{axis}
\end{tikzpicture}